\documentclass{vospwks2007}

\title{Cool objects: From SED fitting to age estimation}
\author[1,2]{A. Bayo}
\author[1,2]{D. Barrado y Navascu\'es}
\author[1]{M. Morales--Calder\'on}
\author[1,2]{E. Solano}
\author[1,2]{C. Rodrigo}
\author[1,2]{R. Guti\'errez}
\author[3]{F. Allard}
\affil[1]{Laboratorio de Astrof\'{\i}sica Espacial y F\'{\i}sica Fundamental (LAEFF-INTA), P.O. 50727, E-28080 Madrid, Spain}
\affil[2]{SVO Thematic Network, Spain}
\affil[3]{Centre de Recherche Astronomique de Lyon (CRAL), Ecole Normale Sup\'erieure de Lyon, 69364, Lyon, France}

\newcommand{\btx}{\textsc{Bib}\TeX}

\begin{document}


\maketitle

\begin{abstract}
The physical properties of almost any kind of astronomical object
can be derived by fitting synthetic spectra or photometry extracted from
theoretical models to observational data.
This process usually involves working with
multiwavelength data, which is one of the cornerstones of
the VO philosophy. From this kind of studies, when
combining with theo\-re\-ti\-cal isochrones one can even
estimate ages. We present here the results obtained from a
code designed to perform $\chi^2$ tests to both spectroscopic
models (or the associated synthetic photometry) and combinations
of blackbodies (including modified blackbodies).
Some steps in this process can already be done in
a VO environment, and the rest are in the process of
development. We must note that this kind of studies in 
star forming regions, clusters, etc. produce a huge amount of data, 
very tedious to analyze using the traditional methodology. This make 
them excellent examples where to apply the VO capabilities.

\end{abstract}

\section{Introduction}

The determination of physical parameters of astronomical objects from observational data
is usually linked with the use of theoretical models as templates.
It is very common to estimate 
gravities of ste\-llar objects by fitting different model/templates to gravity--sensitive lines 
present in observed spectra. Another usual method is to compare slopes in Spectral 
Energy Distributions constructed from observational photometric data with synthetic ones derived
using theoretical spectra.\\
These commonly used methods easily turn to be tedious when being applied to large amount 
of data. Nowadays, astronomers deal with very large databases, not only consisting in observational
data but also models from di\-ffe\-rent groups that need to be combined in order to improve the results.\\
The Virtual Observatory seems to be one of the possible solutions to some intrinsic problems that 
this new way of doing astronomy generates. VO-tools\footnote{\tt http://ivoa.net/twiki/bin/IVOA/IvoaApplications} 
now allow astronomers to deal with their own observational or theoretical databases and compare them 
with others already published in the ``on-line literature''.\\
However, the Virtual Observatory is still under development, and there are some very useful and common
tasks, in particular in the field of spectroscopy, that could be implemented in the near future.

\section{The scientific case}

Our group is involved in a comprehensive study of the Lambda Orionis Star Forming Region. 
This region includes several distinct associations: Collinder 69,
Barnard 30 and 35, as well as other dark clouds, LDN 1588 and 1603.
We use as an example in this work Collinder 69. Although we have collected multiwavelength photometric data 
from the optical to the mid-infrared regime in this association, in
this work we have only used IRAC (the mid-infrared camera on-board the 
Spitzer Space Telescope) photometry \citep{Barrado07}.
Therefore, the original data consists of four photometric points (when detected) per object of the sample.

Our goals are the following:\\
1.- Build the Spectral Energy Distributions for every object in our sample using the photometric data 
available in astronomical archives and services (a typical VO task).\\
2.- Obtain theoretical synthetic spectra corresponding to the regime of physical parameters of interest
(models form the Lyon group \citealt{Allard01}).\\
3.- Compute the synthetic photometry from the theoretical models and the appropriate filter systems 
(mainly DENIS \citealt{DENIS}, 2MASS \citealt{2MASS} and Spitzer/IRAC).\\
4.- Perform independent statistical tests to determine the best fitting model, and the best combination 
of blackbody + modified blackbody (for those objects possibly harboring a disk) for each target. 
With these fittings we can estimate different parameters: T$_{\rm eff}$ and gravity when fitting
theoretical models, and an estimation of the distribution of temperatures of the disk when fitting
a modified blackbody.\\
5.- Compare the physical parameters obtained with theoretical isochrones and evolutionary tracks in order
to have an age and mass estimation per each target of the sample, and an age estimation for the sample 
as a whole.\\
\section{The method}
We have followed the steps stated in the previous section trying to use only VO-compliant tools during the
process. \\
We have found that one of the items (number four) cannot be made in the VO-environment since 
there is no tool that fulfill our requirements. Thus we have developed IDL and PERL codes to complete this 
step.\\
We also find problems in step number two since not all the filter transmission curves were available 
in the VO Filter Profile Service\footnote{\tt http://voservices.net/filter/filterlist.aspx?mode=keyword\&keyword=}. 
In those cases (DENIS and IRAC) we had to query directly the web pages 
of the respective consortia. Another point was that there is no VO-tool that performs the composition
filter + model = synthetic photometry (taking into account the rebinning and normalization issues).
\subsection{Virtual Observatory Tools}

Following the scheme of section 2, we started by visualizing our data with TOPCAT, and ``sending'' this
VOTable to Aladin to query some photometric catalogues (mainly DENIS and 2MASS) looking for counterparts 
in a given radius.\\
We performed a parallel query with VOSED\footnote{\tt http://sdc.laeff.inta.es/vosed/} to look for 
spectroscopic observations of any of our sources available through the different data servers. The
result of this query was that there was not available data.\\
At this point we were able to visualize the constructed SEDs with different tools such as 
VOSpec\footnote{\tt http://esavo.esa.int/vospec}, 
SPECVIEW\footnote{\tt http://www.stsci.edu/resources/software\_hardware/specview/users}
and SPLAT\footnote{\tt http://www.starlink.ac.uk/splat}.\\
The next step was to download the theoretical spectra to be used in our fitting process.
They are accessible in a VO-environment from the Spanish Virtual Observatory Theoretical Model Web Server: 
http://laeff.inta.es/svo/theory/db2vo/html/  \\
Once we had the grid of models, we extracted the synthetic photometry with our codes and
performed $\chi^{2}$ test to determine the best fitting model (and combination of blackbody + modified
blackbody for the objects harboring a disk). This kind of fittings provide us with estimations of
T$_{\rm eff}$ and gravity that we can use to compare with the\-o\-re\-ti\-cal isochrones and evolutionary
tracks (available at: http://laeff.inta.es/svo/theory/draw/getiso.php?inises=guest) and thus to estimate 
ages and masses.

\section{Conclusions}

We have estimated different physical parameters (T$_{\rm eff}$, gravity, mass, age, radius and distance) 
for 170 candidate members of Collinder 69.
To achieve these estimations, we have made use of some of the Virtual Observatory capabilities (such as
queries to different databases, including transmission curve filters, catalogues cross-matches, 
arithmetical o\-pe\-ra\-tions on different columns from catalogues, etc.)

In the process of this work we have identified some tasks that are performed in a non VO-compliant
environment. We think the implementation of these tasks in VO tools could be one of the next steps in 
the development of the Virtual Observatory, since these are very commonly used tasks that involve
interoperability and other VO main bases.

\section*{Acknowledgments}

This research has been funded by Spanish grants MEC/ESP2004-01049, MEC/Consolider-CSD2006-0070,
and CAM/PRICIT-S-0505/ESP/0361, and has made use of the Spanish Virtual Observatory supported from 
the Spanish MEC through grants AyA2005-04286, AyA2005-24102-E.
A. Bayo wishes to acknowledge the Spanish Ministry of Education and Science for the financial 
support of a graduate fellowship.

%\begin{thebibliography}{}

%descomentar cuando tenga la clase
%\bibliographystyle{vospwks2007}
%\bibliography{biblio.bib}

%\end{thebibliography}

\end{document}